\documentclass{pasj00}
\draft

\begin{document}
\SetRunningHead{Miyamoto et al. }{Hot Ammonia in NGC~3079}
\Accepted{}

\title{Hot Ammonia in the Center of the Seyfert~2 galaxy NGC~3079}

\author{Yusuke \textsc{Miyamoto}\thanks{Present Address: Center for Astronomy, Ibaraki University, 2-1-1 Bunkyo, Mito, Ibaraki 310-8512},
Naomasa {\sc Nakai}, Masumichi {\sc Seta}, Dragan {\sc Salak}, 
Kenzaburo {\sc Hagiwara}\thanks{Present Address: NEC Aerospace Systems, Ltd., 1-10, Nissin-cho, Fuchu, Tokyo, 183-8501}, 
Makoto {\sc Nagai} and 
Shun {\sc Ishii}\thanks{Present Address: Institute of Astronomy, School of Science, University of Tokyo, Osawa 2-21-1, Mitaka, Tokyo, 181-0015},
  }
\email{miya@mx.ibaraki.ac.jp, 
nakai@physics.px.tsukuba.ac.jp, 
seta@physics.px.tsukuba.ac.jp, 
salak.dragan.fm@u.tsukuba.ac.jp, 
k-hagiwara@wr.jp.nec.com, 
nagai.makoto.ge@u.tsukuba.ac.jp, 
sishii@ioa.s.u-tokyo.ac.jp}
\affil{Division of Physics, Faculty of Pure and Applied Sciences, University of Tsukuba, 1-1-1 Ten-nodai, Tsukuba, Ibaraki 305-8571}
\and
\author{Aya {\sc Yamauchi}}
\email{a.yamauchi@nao.ac.jp}
\affil{Mizusawa VLBI Observatory, National Astronomical Observatory of Japan, 2-12 Hoshigaoka, Mizusawa, Oshu, Iwate 023-0861}


%


\maketitle

\begin{abstract}
We present the results of ammonia observations toward the center of NGC 3079.
The NH$_3(J,K)=(1,1)$ and $(2,2)$ inversion lines were detected in absorption with the Tsukuba 32-m telescope, and the NH$_3 (1,1)$ through $(6,6)$ lines with the VLA, 
although the profile of NH$_3 (3,3)$ was in emission in contrast to the other transitions.
The background continuum source, whose flux density was $\sim50$~mJy, could not be resolved with the VLA beam of $\lesssim 0\farcs09\times0\farcs08$. 
All ammonia absorption lines have two distinct velocity components: 
one is at the systemic velocity and the other is blueshifted, 
and  both components are aligned along the nuclear jets.
For the systemic components, the relatively low temperature gas is extended more than the high temperature gas. 
The blueshifted  NH$_3 (3,3)$ emission can be regarded as ammonia masers  associated with shocks by strong winds probably from newly formed massive stars or supernova explosions in  dense clouds in the nuclear megamaser disk. 
Using para-NH$_3 (1,1)$, $(2,2)$, $(4,4)$ and $(5,5)$ lines with VLA, we derived the rotational temperature $T_{\rm rot}=120\pm12$~K and $157\pm19$~K for the systemic and blueshifted components, respectively. 
The total column densities of NH$_3(0,0)$--$(6,6)$, assuming $T_{\rm ex}\approx T_{\rm rot}$,  were $(8.85\pm0.70)\times10^{16}$~cm$^{-2}$ and $(4.47\pm0.78)\times10^{16}$~cm$^{-2}$ for the systemic and blueshifted components, respectively.
The fractional abundance of NH$_3$ relative to molecular hydrogen H$_2$ for the systemic and blueshifted was [NH$_3$]/[H$_2]=1.3\times10^{-7}$ and $6.5\times10^{-8}$, respectively.
 We also found the $F=4$--4 and $F=5$--5 doublet lines of OH~$^2\Pi_{3/2}$~$J=9/2$ in absorption, which could be fitted by two velocity components, systemic and redshifted components. 
The rotational temperature of OH was estimated to be $T_{\rm rot, OH}\geq175$~K, 
tracing hot gas associated with the interaction of the fast nuclear outflow with dense molecular material around the nucleus.
\end{abstract}

\section{Introduction}

\begin{figure}
 \begin{center}
  \includegraphics[width=160mm]{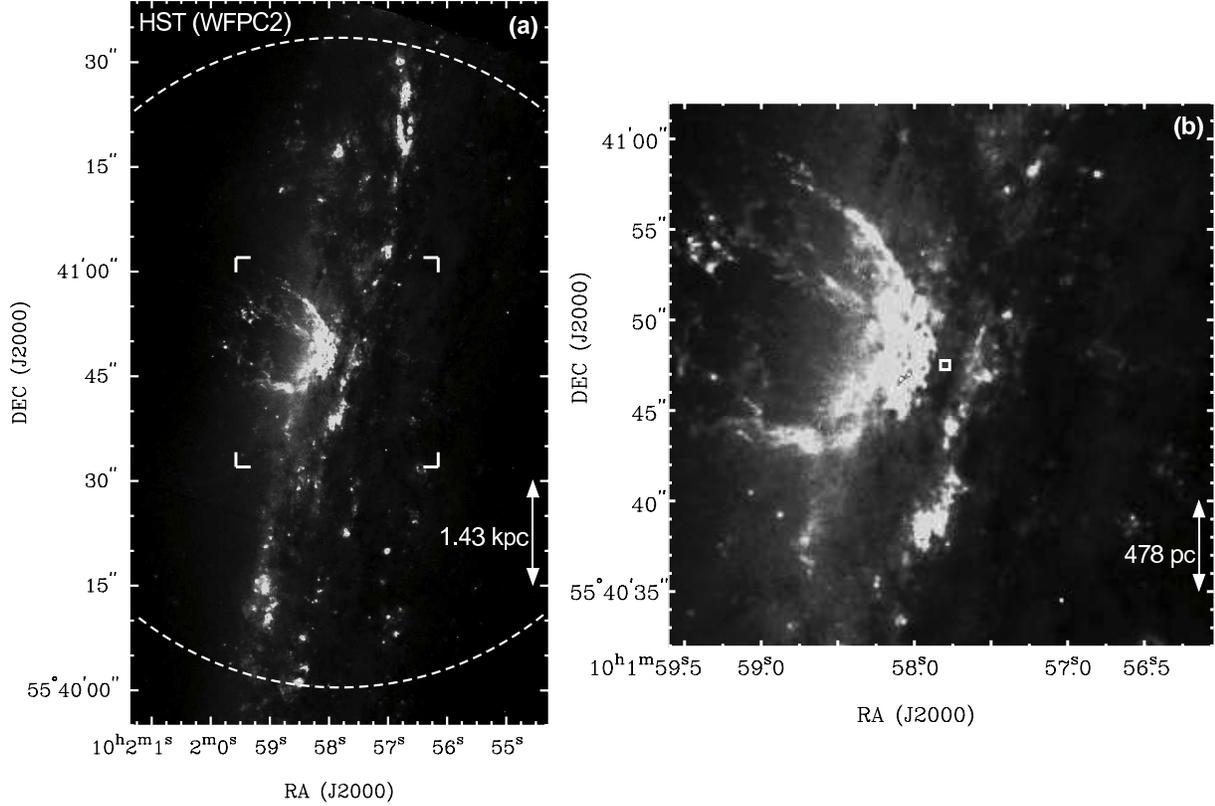} 
 \end{center}
\caption{Optical image of NGC~3079.
(a) {\it HST} WFPC2 image of  [{N}\emissiontype{II}] and H\emissiontype{$\alpha$} line emissions, obtained from the HST data archive.
The dashed lines represent the beam size of ammonia observations with the Tsukuba 32-m ({\it HPBW}$=93\arcsec$.)
(b) Enlargement of the central region in (a).
The region marked by the solid box corresponds to the frame of figure~\ref{fig:map1}.
}
\label{fig:HST}
\end{figure}

Galactic winds influence galaxy evolution because they play an important role in the cycle of material transport in the galaxy. 
The winds are driven by supernovae and/or AGNs. 
One of the clearest examples of a superwind is the prominent bubble emerging from the nucleus (e.g., \cite{duric1988}) of the  edge-on ($i=84\arcdeg$; \cite{irwin1991}) disk galaxy NGC 3079 (figure~\ref{fig:HST}) at a distance of 19.7~Mpc \citep{springob2009}. 
{NGC~3079} is classified as a LINER \citep{heckman1980} or Seyfert 2 galaxy \citep{ford1986}. 
Optical spectroscopy showed gas motions with high velocity across the lobes and unusually the high [{N}\emissiontype{II}]/H\emissiontype{$\alpha$} line ratio  
which indicates the presence of shocks \citep{veilleux1994}. 
\citet{cecil2001} explained the morphology and kinematics of gas in {NGC~3079} by a model of a starburst-driven wind.
\citet{yamagishi2010}, however, reported a relatively low star formation rate of 2.6 M\Sol yr$^{-1}$ for the central 4-kpc region of the galaxy from infrared observations with the AKARI satellite.
\citet{duric1988} proposed an alternative model in which an AGN-driven wind from the nucleus was directed toward the galaxy minor axis by interactions with dense gas surrounding the nucleus. 
The interactions cause shocks that explain the observed strength of H$_2$ $\nu=1$--0 $S(1)$ emission by collisional excitation \citep{hawarden1995, meaburn1998} and heat dust to a temperature of  $\sim1000$~K \citep{israel1998},   
although the contribution of the jets from the AGN to the H$\alpha$ flux of 
the lobes is $\sim10\%$  \citep{cecil2001}. 

Investigations of the physical properties of  molecular gas around the nucleus are helpful 
to understand  the nuclear power source of {NGC~3079}.  
The molecular gas is abundant  in the central region (\cite{sofue1992, irwin1992, sofue2001, koda2002}) and the dense gas  concentrates toward the center \citep{kohno2001}. 
The concentration can be explained by gas inflow toward the center as a result of 
the removal of the gas angular momentum with  bar potential (e.g., \cite{nishiyama2001}), where 
the bar structure in {NGC~3079}  has been resolved clearly in the velocity domain \citep{koda2002}. 

Ammonia (NH$_3$) is a useful thermometer (e.g., \cite{walmsley1983, danby1988}) for relatively dense molecular gas ($n_{\rm H_2} \sim 10^{3-4}$~cm$^{-3}$).
NH$_3$ has a symmetrical top structure which shows inversion doublets caused by the nitrogen atom tunneling through the potential barrier at the plane of three hydrogen atoms. 
The allowed  dipole transitions of NH$_3$ are $\Delta J=\pm1$ and $\Delta K=0$, because the dipole moment corresponds to the symmetry axis of the molecule.  
The non-metastable levels $[J, K(\ne J)]$ decay rapidly (Einstein $A$-coefficients $\sim10^{-1}$~s$^{-1}$) via the far-infrared $\Delta J=1$ transitions, 
and the radiative $\Delta K=\pm3$ transitions are very slow ($A \sim10^{-9}$~s$^{-1}$; \cite{oka1971}), hence 
the metastable $[J, K(= J)]$ levels ($A\sim10^{-7}$~s$^{-1}$) are populated.
The relative populations of the metastable levels are mainly determined by collisions  
and thus follow the Boltzmann distribution.
The rotational temperature can be derived from the ratio of column densities of the metastable levels.
The adjacent inversion lines in frequency  (see subsection~\ref{subsec:tkb_obs}) allow us 
to measure the lines simultaneously with the same telescope and receiver
and thus to evaluate the line ratios accurately owing to the similar beam sizes, same telescope pointing and atmospheric conditions.

In this paper we report the detections of NH$_3$ (plus highly excited OH) lines toward the center of {NGC~3079}. 
We derive their rotational temperatures and estimate the NH$_3$ column density and the abundance under high temperature.
The basic parameters of {NGC~3079} adopted in this paper are summarized in table~\ref{tab:para}. 
Velocities used here are in the radio definition and with respect to the local standard of rest (LSR).

\section{Observations}
\subsection{Tsukuba 32-m}
\label{subsec:tkb_obs}

 \begin{table}
 \begin{center}
\caption{Parameters of {NGC 3079}}\label{tab:para}
  \begin{tabular}{ccc}
   \hline
   \multicolumn{1}{c}{Parameter} & {Value} & {Reference} \\
   \hline
	R.A. (J2000.0) 		& 10$^{\rm h}$01$^{\rm m}$57\fs803 	& 1 \\
	Decl. (J2000.0) 	& +55\arcdeg40\arcmin47\farcs24 		& \\
	Distance 			& 19.7~Mpc 						& 2\\
	Morphological type 	& SB(s)c edge-on 					& 3 \\
	Nuclear Activity 	& LINER / Seyfert 2  				& 4, 5 \\
	Systemic Velocity (LSR) & 1116~km~s$^{-1}$ 				& 6 \\
   \hline\\
   \multicolumn{3}{@{}l@{}}{\hbox to 0pt{\parbox{85mm}{\footnotesize
      References. --- (1) \cite{petrov2011}; (2) \cite{springob2009}; (3) \cite{devaucouleurs1991};
(4) \cite{heckman1980}; (5) \cite{ford1986}; (6) \cite{irwin1991}
     }\hss}}
   \end{tabular}
 \end{center}
 \end{table}

Observations of ammonia toward the center of {NGC 3079} were made in March and April 2008 
with the Tsukuba 32-m telescope of the Geospatial Information Authority of Japan.
The full half-power beam width was HPBW$=93\arcsec\pm6\arcsec$ at 24~GHz, 
 corresponding to $8.9$~kpc at the distance of the galaxy (19.7~Mpc).
The main beam efficiency of the antenna  $\eta_{\rm mb}$ was measured, using Jupiter whose brightness temperature was adopted to be $T_{\rm b}=138\pm7$~K at $\lambda=1.3$~cm \citep{depater2005},
and depended on the elevation of the antenna with the maximum value of $\eta_{\rm mb}=0.44\pm0.02$ at $EL=38\arcdeg$ at 24~GHz. 
The aperture efficiency was maximumly $\eta_{\rm a}=0.37\pm0.01$ at $EL=38\arcdeg$, 
calculated from the main beam efficiency and the beam size, 
and confirmed by observing {3C~286}.
The sensitivity of the antenna was $S/T_{\rm A}^{\ast}=9.28$~Jy~K$^{-1}$ at 24~GHz, where $S$ is the flux density and $T_{\rm A}^{\ast}$ the antenna temperature.

The receiver front-end utilized a HEMT amplifier cooled to 11~K, equipped with a circular polarized feed.
Its frequency coverage was 19.5--25.1~GHz.
The receiver back-end was a 16384-channels FFT spectrometer.
The total bandwidth and frequency resolution of the spectrometer were about 1.0~GHz and $55$~kHz, 
which corresponded to $1.3\times10^4$~km~s$^{-1}$ and $0.69$~km~s$^{-1}$ at 24~GHz, respectively. 
The bandwidth allowed us to observe four inversion transitions of ammonia, $(J,K)=(1,1)$--$(4,4)$, simultaneously (see table~\ref{tab:obs1}).

The line intensity was calibrated by the chopper wheel method, 
yielding an antenna temperature, $T_{\rm A}^{\ast}$, corrected for both atmospheric and antenna ohmic losses \citep{ulich1976}. 
The typical system noise temperatures during the observations were 80-120~K (SSB) in $T_{\rm A}^{\ast}$ at observing elevations. 
The main beam brightness temperature, $T_{\rm mb} (\equiv T_{\rm A}^{\ast}/\eta_{\rm mb})$, was converted from $T_{\rm A}^{\ast}$  using the main beam efficiency of the antenna at each observing elevation.
The observations were made by position-switching with an integration time of 10~s per scan.
The telescope pointing was checked every 1~hr by observing H$_2$O maser of the Mira-type variable  star {R UMa}, and the resultant pointing error was about $10\arcsec$--$20\arcsec$.

The spectral data were reduced with the NEWSTAR software which was developed at the Nobeyama Radio Observatory\footnotemark (NRO).
\footnotetext{The Nobeyama Radio Observatory is a branch of the National Astronomical observatory of Japan, National Institute of Natural Sciences.}
The spectra were flagged and averaged after linear baseline subtraction, resulting in total observing time (ON source) was about 10.5 hours.
All of the averaged spectra were bound up every 64 channels to reduce the noise level, 
resulting in a frequency resolution of 3.6~MHz or a velocity resolution of 45~km~s$^{-1}$ at 24~GHz.

\subsection{VLA}
Observations of ammonia in {NGC 3079} were carried out
with the A-configuration of the Karl G. Jansky Very Large Array (VLA) of the National Radio Astronomy Observatory\footnotemark (NRAO) of USA over a period from October 2012 through January 2013.
The K-band receivers and the new WIDAR correlators were used for the dual-polarization.
\footnotetext{The National Radio Astronomy Observatory is a facility of the National Science Foundation operated under cooperative agreement by Associated Universities, Inc.}
The correlator  in each polarization was composed of 8 subbands whose bandwidth and frequency resolution were 128~MHz and 1~MHz, respectively, 
and thus the total bandwidth was 1024~MHz.
The total bandwidth allowed us to observe NH$_3 (J,K)=(1,1)$--$(3,3)$ simultaneously in the frequency band centered on 23.525~GHz and later $(J,K)=(4,4)$--$(6,6)$ centered on 24.590~GHz.
The different UV coverages for the low and high frequency setting resulted in the different sizes of the synthesized beams  (table~\ref{tab:obs1}).
Observing time was  5~hrs for each frequency band, and  totally 10~hrs. 
Antenna pointing was checked every 60 min by observing {J0958+6533} with the X-band receiver, 
and the flux density at the K-band was calibrated by {3C 147}.
We tracked {J0958+6533}  every 2.5~min to calibrate time variations of amplitude and phase. 
The data were also used to determine the bandpass.

The observed data were processed using the Common Astronomy Software Applications (CASA; \cite{mcmullin2007}).
The data obtained at different observing tracks were combined 
after subtracting continuum emission determined at the absorption-free channels and rearranging the velocity resolution to be 15~km~s$^{-1}$ and the velocity range to be 550--1750~km~s$^{-1}$.
To image the continuum emission, 
we used the flux density at the absorption-free channels.
The imaging was performed with the CLEAN-algorithm in CASA.
CLEAN maps were obtained considering the briggs weighting mode on the data with robust of 0.5. 
The resultant maps are $300\times300$~pixels with $0\farcs01$~per pixel. 
The synthesized beams are given in table~\ref{tab:obs1}.

 \begin{table}
 \begin{center}
\caption{Observational parameters with the VLA}\label{tab:obs1}
  \begin{tabular}{ccc}
   \hline
   \multicolumn{1}{c}{Transitions} & {Frequency (GHz)} & {Beam Size (arcsec)\footnotemark[$\ast$]}  \\
   \hline
NH$_3(2,1)$	&	23.098819		&	$0.083\times0.077$	\\
NH$_3(1,1)$	&	23.694506		&	$0.083\times0.077$	\\
NH$_3(2,2)$	&	23.722634		&	$0.083\times0.077$	 \\
NH$_3(3,3)$	&	23.870130		&	$0.083\times0.077$	 \\
NH$_3(4,4)$	&	24.139417		&	$0.091\times0.071$	 \\
NH$_3(5,5)$	&	24.532989		&	$0.089\times0.069$	 \\
NH$_3(6,6)$	&	25.056025		&	$0.087\times0.067$	 \\
OH $^2\Pi_{2/3} $$J=9/2$ $F=4-4$	&	23.8176153	&	$0.083\times0.077$	\\
OH $^2\Pi_{2/3} $$J=9/2$ $F=5-5$	&	23.8266211	&	$0.083\times0.077$	\\
   \hline\\
   \multicolumn{3}{@{}l@{}}{\hbox to 0pt{\parbox{85mm}{\footnotesize
 \footnotemark[$\ast$] 0\farcs08 corresponds to 7.6~pc at the distance of {NGC~3079}.
     }\hss}}
   \end{tabular}
 \end{center}
 \end{table}


\section{Results}
\subsection{Tsukuba 32-m}
\begin{figure}
 \begin{center}
  \includegraphics[width=8cm]{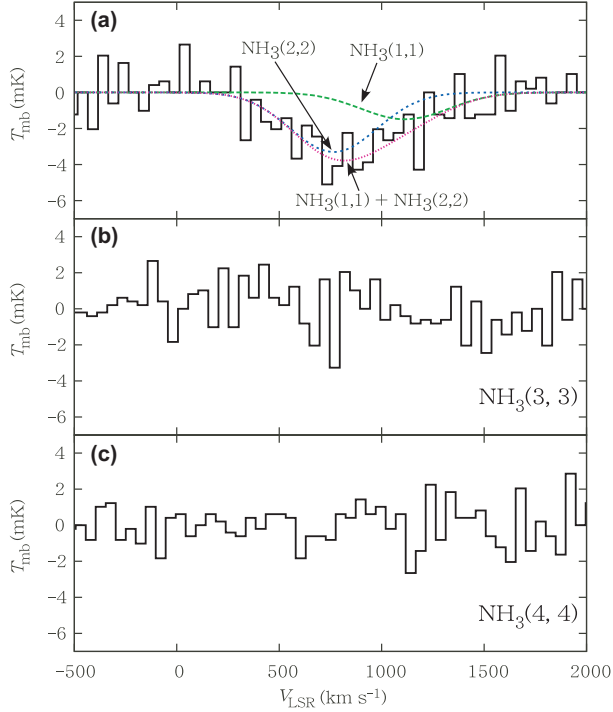} 
 \end{center}
\caption{Continuum-subtracted spectra of NH$_3$ toward the central region of NGC 3079 with the Tsukuba 32-m. 
The velocities $V_{\rm LSR}$ are with respect to the rest frequencies of NH$_3 (1,1)$, NH$_3(3,3)$, and NH$_3(4,4)$ in figure (a), (b), and (c), respectively.}
\label{fig:32m}
\end{figure}

Absorption lines of NH$_3$ $(J,K)=(1,1)$ and $(2,2)$  toward the center of {NGC 3079} were detected with the Tsukuba 32-m telescope (figure~\ref{fig:32m}).
The two lines are so broad that they overlap each other for the rest frequency difference of 28.1~MHz corresponding to the velocity difference  $\Delta V=356$~km~s$^{-1}$.
To separate the spectrum into NH$_3 (1,1)$ and $(2,2)$ components,
we adopt a double Gaussian fitting function, 
\begin{eqnarray}
F(v)=T_{\rm mb, 11}\exp \frac{ -(v-V_{\rm LSR, 11})^2}{2\sigma_{11}^2} +T_{\rm mb, 22}\exp\frac{ -(v-V_{\rm LSR, 22}-\Delta V)^2}{2\sigma_{22}^2 } ,
\end{eqnarray}
where $v$ is $V_{\rm LSR}$ in figure~\ref{fig:32m}, $T_{\rm mb, 11}$ and $T_{\rm mb, 22}$  the main-beam brightness temperature, 
$V_{\rm LSR, 11}$ and $V_{\rm LSR, 22}$ the velocity relative to the local standard of rest (LSR), 
and $\sigma_{11}$ and $\sigma_{22}$ the velocity dispersion of NH$_3(1,1)$ and $(2,2)$, respectively.
Figure~\ref{fig:32m} shows the result of least-square fitting of the function $F(v)$ to the observed spectrum,  where the green, blue, and red lines show NH$_3(1,1)$, $(2,2)$, and the combination of them, respectively.
Table~\ref{tab:fit1} gives the fitting parameters of the main-beam brightness temperatures ($T_{\rm mb}$), the velocities ($V_{\rm LSR}$), and the full-width at half-maximum ($\Delta v_{1/2} =\sqrt[]{\mathstrut 8\ln 2}\sigma$).
The  velocities, $V_{\rm LSR}=1113\pm83$~km~s$^{-1}$ for NH$_3(1,1)$ and $1122\pm38$~km~s$^{-1}$ for NH$_3(2,2)$ are consistent with the systemic velocity of the galaxy $V_{\rm sys}=1116\pm1$~km~s$^{-1}$  \citep{irwin1991},  
within the errors.
NH$_3(3,3)$ line is undetected at the noise level of $\Delta T_{\rm mb}\sim 1.39$~mK as observations with the GBT \citep{mangum2013}. 
In addition, the noise level of $\Delta T_{\rm mb}\sim 1.23$~mK is not low enough to detect an absorption line of NH$_3(4,4)$  whose intensity is expected to be comparable to that of NH$_3(1,1)$. 

The peak flux densities for NH$_3(1,1)$  and NH$_3(2,2)$ are derived to be $-5.7\pm1.8$~mJy  and $-13.2\pm2.0$~mJy, respectively, from the conversion of $S/T_{\rm A}^{\ast}$ in section~\ref{subsec:tkb_obs}.
These flux densities are lower than those with the GBT ($-18.5$~mJy and $-22.6$~mJy for NH$_3(1,1)$ and NH$_3(2,2)$, respectively; \cite{mangum2013}).
Since the beam sizes (HPBW $\approx 93\arcsec$) of the Tsukuba 32-m is larger than those (HPBW $\approx 30\arcsec$) of GBT,
the difference may be caused (1)  by the beam dilution of the absorption lines, and
(2) by contamination from emission in the galactic disk 
because the molecular gas is extended over the radius of $r\sim70\arcsec$ \citep{braine1997} while 
it is expected that the continuum source is  over $r\sim15\arcsec$ (e.g., \cite{duric1988}).
We, however, should mention the possibility that the spectra for the Tsukuba 32-m are affected by instrumental and atmospheric instabilities that lead to poor spectral baselines and uncertain line profiles.

 \begin{table}
 \begin{center}
	\caption{The fitted parameters}
	\label{tab:fit1}
		\begin{tabular}{cccc}
		\hline
  		 \multicolumn{1}{c}{Transitions} & {$T_{\rm mb}$ (mK)} & {$V_{\rm sys}$ (km~s$^{-1}$)} & {$\Delta v_{1/2}$ (km~s$^{-1}$)} \\
		 \hline
		NH$_3(1,1)$	&	$-1.43\pm0.45$	&	$1113\pm83$	&	$504\pm167$ \\
		NH$_3(2,2)$	&	$-3.31\pm0.47$	&	$1122\pm38$	&	$499\pm75$ \\
   \hline
   \end{tabular}
 \end{center}
 \end{table}

\subsection{VLA}
\subsubsection{Continuum Emission}
We detected an unresolved continuum source (contours in figure~\ref{fig:map1}) with the resolution of $\leq0\farcs09\times0\farcs08$ (beam size in table~\ref{tab:obs1}), corresponding to $8.6\times7.6$~pc, at R.A.~(J2000.0)$=10^{\rm h} 01^{\rm m} 57\fs803$, Decl.~(J2000.0)$=55$\arcdeg40\arcmin47\farcs25,  
which was the same as the position measured with VLBI at 5~GHz, R.A.~(J2000.0)$=10^{\rm h} 01^{\rm m} 57\fs8034$, Decl.~(J2000.0)$=55$\arcdeg40\arcmin47\farcs243 \citep{petrov2011}. 
The measured flux density of about $50$~mJy (table~\ref{tab:fit_nh3}) is smaller than the value of $\sim80$~mJy at $\lambda\sim1.2$~cm expected from the flux density of 65~mJy at 4866~MHz and the spectrum index of $\alpha=+0.15$ (where $S_{\nu}\propto \nu^{\alpha}$) determined from the ratio of 1465 and 4866 MHz flux densities with the angular resolution of $\sim1\arcsec$ \citep{duric1988},
and than the value of 174~mJy at $\lambda\sim1.2$~cm  measured with the beam size of $\sim30\arcsec$ by  the GBT  \citep{mangum2013}.
On the other hand, within a 0\farcs015 radius from the center, 
some distinct continuum sources were identified with VLBI, 
and the total flux density was $38\pm5$~mJy at $\lambda\sim1.3$~cm
[components A, B, C and E in \citet{yamauchi2004}].
Considering these results and the detectable largest angular scale of $\sim 2\arcsec$  by the VLA,  
it is indicated that there are continuum sources extended to more than $r\sim1\arcsec$. 

\subsubsection{Ammonia Inversion Lines}
\label{subsec:ammonia}
\begin{figure}
 \begin{center}
  \includegraphics[width=10cm]{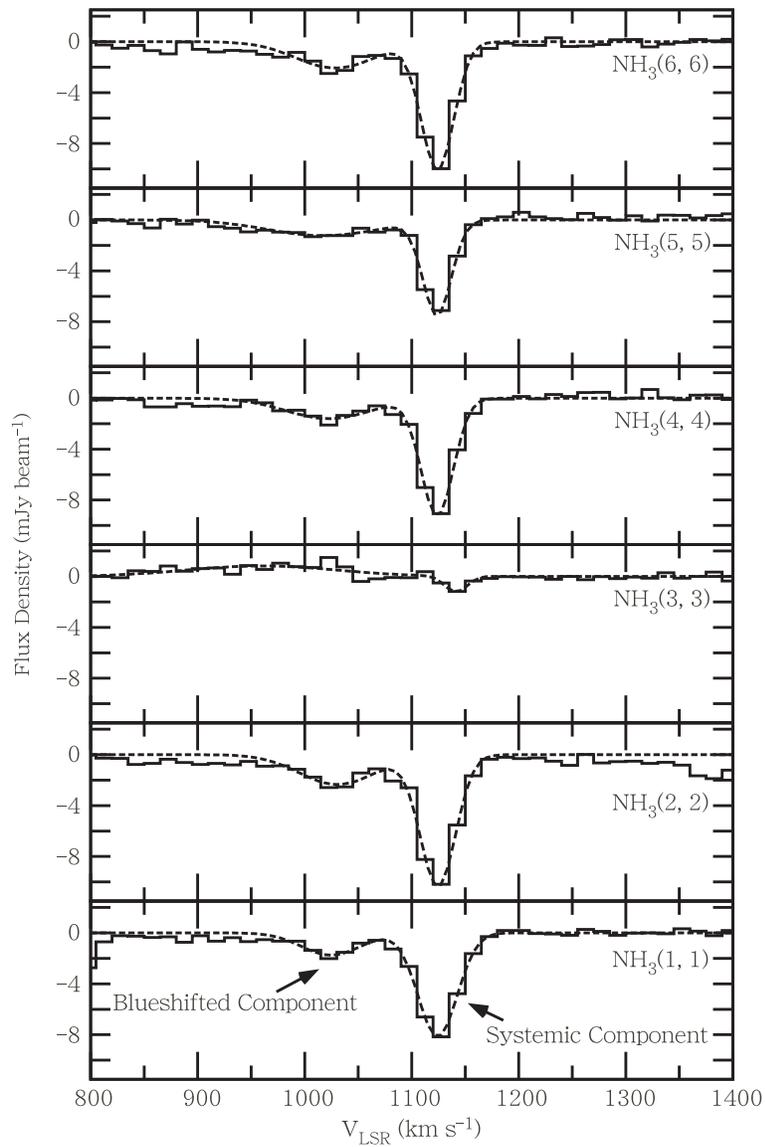} 
 \end{center}
\caption{Continuum-subtracted NH$_3$ lines at the center of {NGC 3079} with the VLA. 
The spectra of NH$_3(J,K)=(1,1)$, ($2,2$), ($3,3$), ($4,4$), ($5,5$), and ($6,6$)  from bottom to top.
Gaussian fits to the lines are overlaid with dotted lines.} 
\label{fig:spe-all}
\end{figure}

Figure~\ref{fig:spe-all} shows the spectra measured with the VLA. 
The six ammonia inversion lines, NH$_3(1,1)$ to $(6,6)$, are detected in absorption against the continuum source,  
although the profile of the NH$_3(3,3)$ line is anomalous.
The spectra of NH$_3(1,1)$ and $(2,2)$ do not overlap each other, rather clearly separate in velocity 
owing to narrower velocity widths, unlike the result of the 32-m (figure~\ref{fig:32m}).
This could be because VLA,  
with synthesized beam size $\approx0\farcs08$ and 
 detectable largest angular scale  $\sim2\arcsec$,   
observed much smaller regions than the 32-m (beam sizes $\approx93\arcsec$) did and thus did not detect components broad in space and velocity (e.g., the rotation velocity of the galactic disk is $\sim260$--$300$~km~s$^{-1}$ at $r\gtrsim1\arcsec$; \cite{sofue2001}).
In figure~\ref{fig:spe-all}, 
all the metastable NH$_3$ absorption lines show two distinct velocity components: 
one is at the systemic velocity ($V_{\rm LSR} \sim 1116$~km~s$^{-1}$)   
and the other is blueshifted ($V_{\rm LSR} \sim 1020$~km~s$^{-1}$). 
We fitted Gaussian profiles to the observed absorption lines.
Table~\ref{tab:fit_nh3} gives 
the resultant parameters of the flux densities ($S$), the velocities ($V_{\rm LSR}$), and  the  optical depths ($\tau$, see equation~(\ref{eq:tau})) at the absorption peak, the line widths ($\Delta v_{1/2}$) and the  optical depths integrated over the line profiles ($\int\tau dv$) for  the systemic and  blueshifted components.
 
\begin{figure}
 \begin{center}
  \includegraphics[width=140mm]{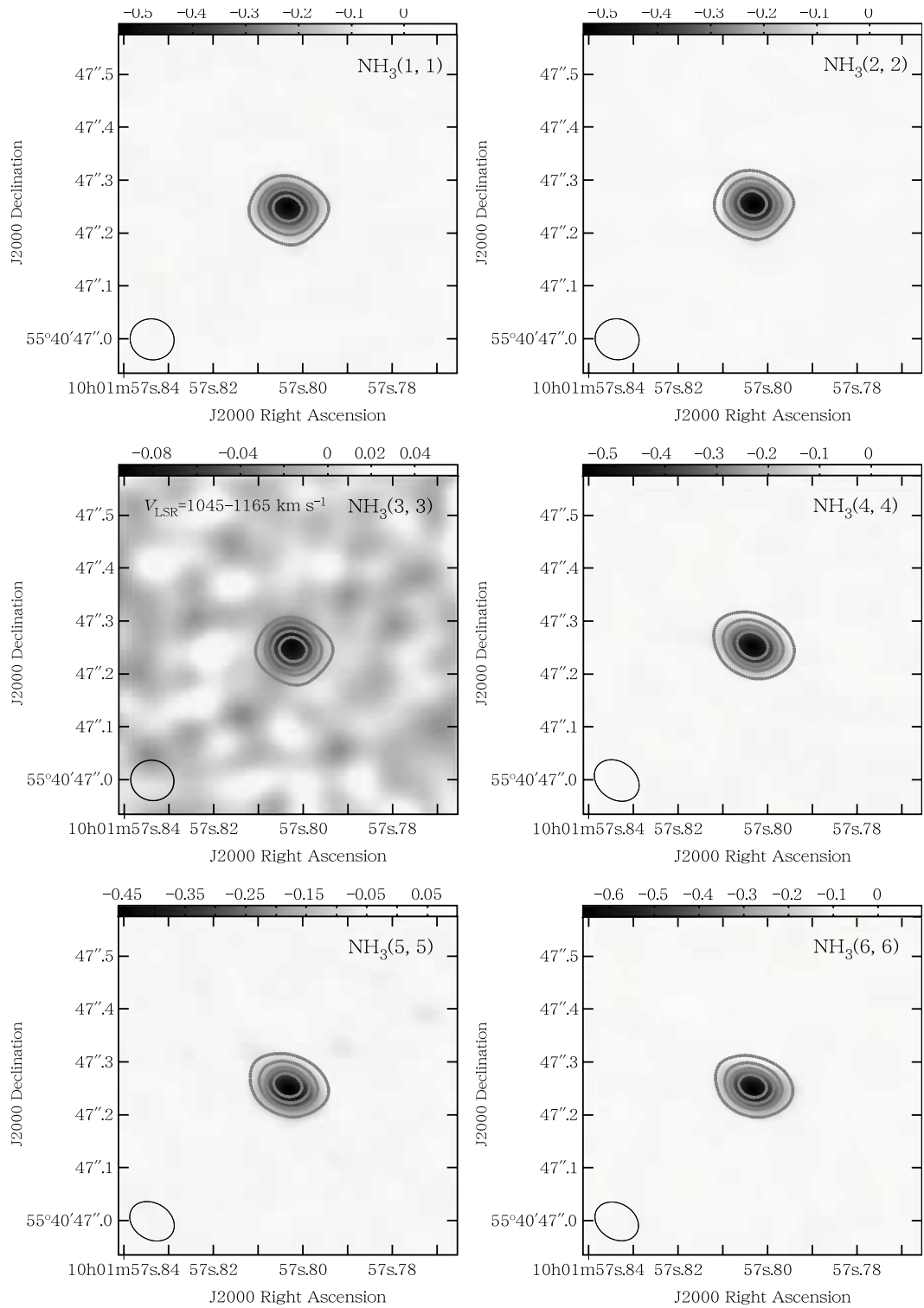} 
 \end{center}
\caption{The continuum map (contours) extracted from the absorption-free regions in band is superposed on the integrated intensity map (Jy~beam$^{-1}$~km~s$^{-1}$) of NH$_3(1,1)$ to (6,6) absorption lines (color). 
The integrated velocity ranges of NH$_3(1,1)$, $(2,2)$, $(4,4)$, $(5,5)$ and $(6,6)$ lines are $V_{\rm LSR}=835$--$1180$~km~s$^{-1}$, and that of NH$_3(3,3)$ is $V_{\rm LSR}=1045$--$1165$~km~s$^{-1}$.
The contours are plotted at 10, 20, 30 and 40 mJy beam$^{-1}$.
The synthesized beam is plotted at the lower left corner.}
\label{fig:map1}
\end{figure}

Figure~\ref{fig:map1} shows the distribution of the intensities of the NH$_3(1,1)$, $(2,2)$, $(4,4)$, $(5,5)$, and $(6,6)$ lines integrated at $V_{\rm LSR}\approx835$--1180~km~s$^{-1}$, 
including both the systemic and blueshifted components,
and of the intensity of the NH$_3(3,3)$ absorption line integrated at $V_{\rm LSR}=$1045--1165~km~s$^{-1}$.
The figure indicates that there is abundant molecular gas in front of the nuclear radio continuum source, 
although the distribution could not be resolved with the spatial resolution of $\lesssim9$~pc (but see section~\ref{subsec:distribution} for more detail discussion). 

The absolute flux densities of NH$_3(1,1)$--$(6,6)$ absorption lines except NH$_3(3,3)$ are about half of those with the GBT whose beam size  is ${\rm HPBW}\approx30\arcsec$ \citep{mangum2013}.
For the systemic components, the ratio of the flux densities with GBT to VLA decreases with the excitation level (table~\ref{tab:ratio}). 
This indicates that the relatively low temperature gas is more extended than the high temperature gas. 
In contrast, for the blueshifted components, 
the flux density ratio increases at the higher transitions of NH$_3(4,4)$ and $(5,5)$.

\begin{figure}
 \begin{center}
  \includegraphics[width=8cm]{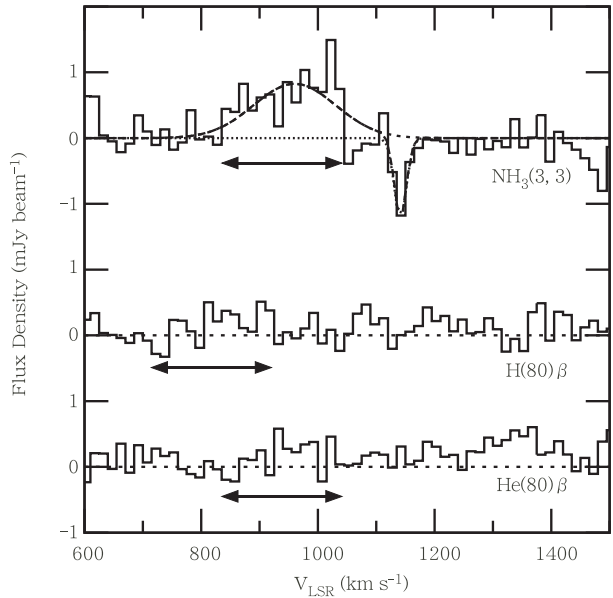} 
 \end{center}
\caption{Enlargement of the NH$_3(3,3)$ spectrum in figure~\ref{fig:spe-all}, and H(80)$\beta$ and He(80)$\beta$ spectra for comparison are shown.
If the emission line in the frame of NH$_3(3,3)$ is attributed to the H(81)$\beta$ or He(81)$\beta$ line,  
the H(80)$\beta$ or He(80)$\beta$ line should be appeared in the range shown by the horizontal arrows, 
but no significant emission can be seen, indicating that the emission at $V_{\rm LSR}\approx830$--1050~km~s$^{-1}$ is not H(81)$\beta$ and He(81)$\beta$ but NH$_3(3,3)$.}
\label{fig:nh3_33}
\end{figure}

At the blueshifted velocity, the NH$_3 (3,3)$ line shows a conspicuous spectrum which is not absorption but emission in contrast to the other NH$_3$ transition (figure~\ref{fig:nh3_33}).
H$(81)\beta$ and He$(81)\beta$ recombination lines, whose frequencies are close to NH$_3(3,3)$ (table~\ref{tab:nh3}), may also explain the emission features.
However, the emissions in the next transition, i.e., H$(80)\beta$ and He$(80)\beta$, were not detected (figure~\ref{fig:nh3_33}).
We therefore focus on  NH$_3 (3,3)$ maser as the most likely candidate of the emission features and discuss in more detail in subsection~\ref{sec:nh3_maser}. 

\subsubsection{OH $^2\Pi_{2/3} J=9/2$ Absorption Lines}
\label{subsec:oh}
We also found a wide absorption line with the width of $\Delta v_{1/2}\gtrsim 290$~km~s$^{-1}$ (figure~\ref{fig:oh}) in the correlator subband which covered  frequencies from 23.781 to 23.909~GHz, including the NH$_3(3,3)$ line.  
Considering the systemic velocity of {NGC 3079} ($V_{\rm sys}=1116$~km~s$^{-1}$), 
the $F=4$--4 and $F=5$--5 doublet lines of OH $^2\Pi_{3/2}$ $J=9/2$ and the HC$_9$N ($J=41$--40) whose rest frequency of 23.8176153, 23.8266211 and 23.822265~GHz\footnote{F. J. Lovas et al., NIST Recommended Rest Frequencies for Observed Interstellar Molecular Microwave Transitions (http://physics.nist.gov/cgi-bin/micro/table5/start.pl).}, respectively, are appropriate candidates for the absorption line.
We, however, ruled out HC$_9$N ($J=41$--40)  
because no absorption feature of the HC$_9$N ($J=40$--39) transition (rest frequency,  23.241246~GHz) was confirmed in our VLA data.
On the other hand, 
detections of the absorption lines at the ground state of OH $^2\Pi_{3/2} J=3/2$ (rest frequencies, 1665- and 1667-MHz) toward the center of NGC 3079  have been reported (e.g., \cite{hagiwara2004}). 
So far the absorption lines at the $^2\Pi_{3/2}$ $J=9/2$ state (excitation level of 511~K above the ground state, \cite{walmsley1986}) 
have been found toward some  compact H\emissiontype{II} regions in the Galaxy (e.g., \cite{winnberg1978,baudry1981,walmsley1986}) and toward the center of {Arp 220} \citep{ott2011}.
We therefore deduce that the absorption is caused by OH $^2\Pi_{3/2}$ $J=9/2$  $F=4$--4 and $F=5$--5. 

\begin{figure}
 \begin{center}
  \includegraphics[width=140mm]{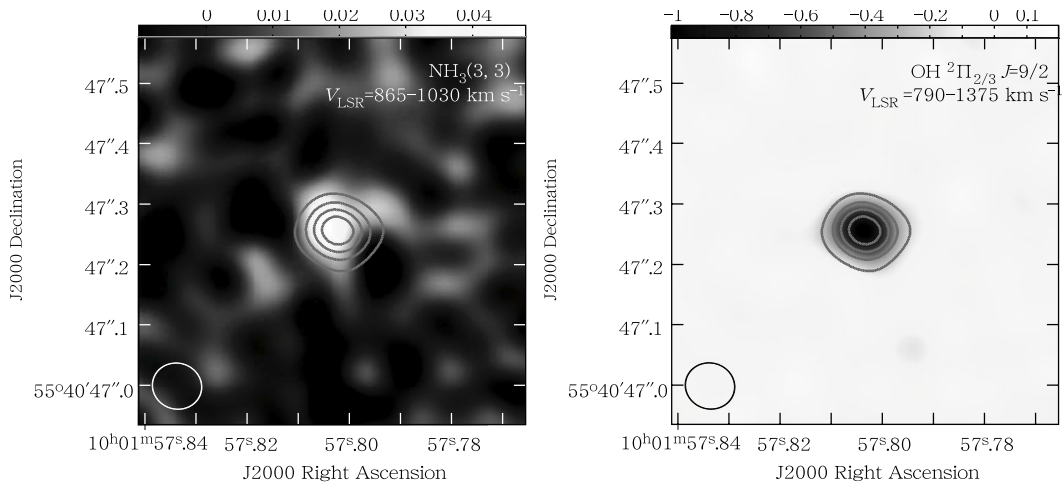} 
 \end{center}
\caption{The integrated intensity map of the blueshifted features of NH$_3(3,3)$ (left) and the OH~$^2\Pi_{3/2}$~$J=9/2$ absorption line (right).
The continuum map (contours) is the same as that of the systemic features of NH$_3(3,3)$ in figure~\ref{fig:map1}.}
\label{fig:map2}
\end{figure}

\begin{figure}
 \begin{center}
  \includegraphics[width=80mm]{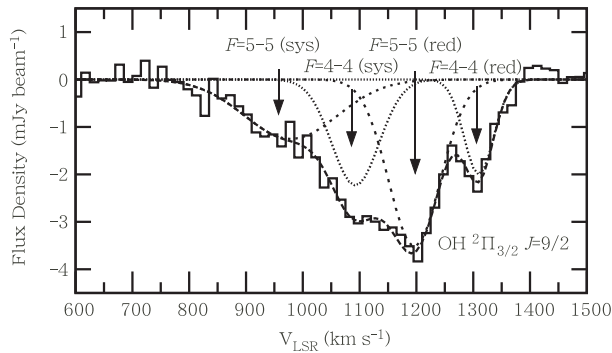} 
 \end{center}
\caption{The absorption spectrum of OH $^2\Pi_{3/2} J=9/2$. 
Gaussian fits to the lines are overlaid.
The velocity is referenced to $\nu=23.8176153$ of OH $^2\Pi_{3/2} J=9/2$ $F=4-4$.}
 \label{fig:oh}
\end{figure}

Figure~\ref{fig:map2}~(right) shows the OH intensity map integrated at $V_{\rm LSR}=790$--1375~km~s$^{-1}$  with respect to the rest frequency of OH~$^2\Pi_{3/2}$~$J=9/2$~$F=4$--4. 
The structure cannot be resolved with the angular resolution of 0\farcs083$\times$0\farcs077. 
The $F=4$--4 and $F=5$--5 doublet lines are spaced  by 113~km~s$^{-1}$, corresponding to the velocity separation between the dips in the wide spectrum, e.g., the components at $V_{\rm sys}\sim1200$~km~s$^{-1}$  and $\sim1310$~km~s$^{-1}$ in figure~\ref{fig:oh}.
Assuming that each doublet line has same velocity components, 
the wide absorption features can be fitted by two velocity components, 
$V_{(\rm OH, sys)}=1085\pm4$~km~s$^{-1}$ and $V_{(\rm OH, red)}=1302\pm2$~km~s$^{-1}$ (table~\ref{tab:fit_oh}).
These velocity components are  different from the velocities at the ground state of OH $^2\Pi_{3/2} J=3/2$ ($1011.9\pm0.9$~km~s$^{-1}$ and $1113.5 \pm 2.0$~km~s$^{-1}$; \cite{hagiwara2004}) and  NH$_3$, 
not showing the blueshifted components but the redshift.
Although {H}\emissiontype{I} absorption  toward the nucleus has the redshifted component \citep{sawada2001}, 
the velocity of $V_{\rm LSR}=1238\pm3$~km~s$^{-1}$ and the line width of $\Delta v_{1/2}= 32\pm6$~km~s$^{-1}$ are inconsistent with those of OH $^2\Pi_{3/2}$ $J=9/2$. 
The highly excited OH line (energy level $=511$~K) therefore traces hot gas with the anomalous dynamics in the central region,   
which may be associated with the hot dust ($T_{\rm d}\sim1000$~K; \cite{israel1998}) heated by the interaction of the fast nuclear outflow with dense and dusty molecular material around the nucleus (e.g., \cite{hawarden1995, meaburn1998}).  
In addition, the wide line widths in  OH $^2\Pi_{3/2}$~$J=9/2$ might be caused by summing many different velocity components around the center, whose distributions are unresolved with the beam size of $\sim 0.08\arcsec$. 
The merged spectra could lead to the widely different ratios of the optical depth of the systemic components to redshifted, e.g.,  1.12 for $F=4-4$ and 0.34 for $F=5-5$. 
To investigate the line ratio in more detail, observations with higher angular resolution enough to resolve the distribution of the OH $^2\Pi_{3/2}$~$J=9/2$ lines would be needed.

 \begin{table}
 \begin{center}
\caption{NH$_3$ parameters}
\label{tab:fit_nh3}
   \begin{tabular}{cccccccc}
   \hline
   \multicolumn{1}{c}{Transitions}		&	{$S$\footnotemark[$\ast$]}	&	{$V_{\rm LSR}$ \footnotemark[$\ast$]}	&	{$\Delta v_{1/2}$}	&	{$\tau$\footnotemark[$\ast$]}	&	{$\int\tau dv$}	&		&	{$S_{\rm cont}$}	\\
{}	&	{(mJy)}	&	{(km~s$^{-1}$)}	&	{(km~s$^{-1}$)}	&		&	{(km~s$^{-1}$)}	&		&	{ (mJy)}\\
   \hline
{NH$_3$}	&	\multicolumn{5}{c}{Systemic Components}	&	{}	&	{}	\\
\cline{2-6} 
$(1,1)$	&	$-8.15\pm0.22$		&	$1117\pm1$	&	$43\pm1$	&	$0.176\pm0.005$	&	$8.01\pm0.35$	&	&	$50.62 \pm0.22$\\
$(2,2)$	&	$-10.28\pm0.38$	&	$1117\pm1$	&	$39\pm2$	&	$0.227\pm0.010$	&	$9.38\pm0.59$	&	&	$50.62 \pm0.30$\\
$(3,3)$	&	$-1.23\pm0.38$		&	$1133\pm4$	&	$36\pm7$	&	$0.025\pm0.008$	&	$0.66\pm0.30$	&	&	$49.99 \pm0.28$\\
$(4,4)$	&	$-9.18\pm0.31$		&	$1116\pm1$	&	$35\pm1$	&	$0.194\pm0.007$	&	$7.30\pm0.41$	&	&	$52.08 \pm0.42$\\
$(5,5)$	&	$-7.34\pm0.28$		&	$1116\pm1$	&	$31\pm2$	&	$0.158\pm0.007$	&	$5.29\pm0.34$	&	&	$50.10 \pm0.39$\\
$(6,6)$	&	$-10.00\pm0.36$	&	$1117\pm1$	&	$36\pm2$	&	$0.228\pm0.010$	&	$8.69\pm0.54$	&	&	$49.09 \pm0.47$\\\\
{}	&	\multicolumn{5}{c}{Blueshifted Components}	&	{}	&	{}	\\
\cline{2-6} 
$(1,1)$	&	$-1.75\pm0.18$		&	$1017\pm3$	&	$65\pm8$ 	&	$0.035\pm0.004$	&	$2.43\pm0.40$	&	&	\\
$(2,2)$	&	$-2.34\pm0.27$		&	$1023\pm5$	&	$84\pm14$	&	$0.047\pm0.006$	&	$4.25\pm0.88$	&	&	\\
$(3,3)$	&	$1.54\pm0.45$		&	$953\pm16$	&	$165\pm40$	&	...			&	...		&	&	\\
$(4,4)$	&	$-1.60\pm0.20$		&	$1015\pm6$	&	$95\pm20$	&	$0.031\pm0.004$	&	$3.14\pm0.77$	&	&	\\
$(5,5)$	&	$-1.23\pm0.15$		&	$1009\pm10$	&	$122\pm34$	&	$0.025\pm0.003$	&	$3.23\pm0.99$	&	&	\\
$(6,6)$	&	$-2.07\pm0.24$		&	$1021\pm5$	&	$87\pm15$	&	$0.043\pm0.005$	&	$4.00\pm0.85$	&	&	\\
   \hline\\
   \multicolumn{8}{@{}l@{}}{\hbox to 0pt{\parbox{85mm}{\footnotesize
 \footnotemark[$\ast$] Values at the absorption peak after Gaussian fitting.
     }\hss}}
   \end{tabular}
 \end{center}
 \end{table}

 \begin{table}
\caption{The observed flux densities of NH$_3$ $(1,1)$$-$$(6,6)$ and 24-GHz continuum with VLA (this work) and GBT. Values in parentheses are standard deviations.}
 \begin{center}
 \label{tab:ratio}
   \begin{tabular}{ccccccccc}
   \hline
	{}	&	\multicolumn{6}{c}{NH$_3$}&{}\\
   \multicolumn{1}{c}{Telescope} & {$(1,1)$} & {$(2,2)$}	&	{$(3,3)$}	&{$(4,4)$} & {$(5,5)$} & {$(6,6)$}	&	{Continuum}\\
   \hline
	{}	&	\multicolumn{6}{c}{Systemic Components~(mJy)}&{(mJy)}\\
	\cline{2-7}
	VLA	&	$-8.15(0.22)$	&	$-10.28(0.38)$	&	$-1.23(0.38)$	&	$-9.18(0.31)$	&	$-7.34(0.28)$	&	$-10.00(0.36)$	&	50\\
	GBT\footnotemark[$\ast$]		&	$-18.51(0.64)$	&	$-22.55(0.59)$	&	...	&	$-19.47(0.69)$	&	$-11.97(0.69)$	&	$-13.67(0.96)$	&	174\\
	$|S_{\rm GBT}$/$S_{\rm VLA}|$ &	$2.27(0.10)$	&	$2.19(0.10)$	&	...	&	$2.12(0.10)$	&	$1.63(0.11)$	&	$1.37(0.11)$	&	3.5\\
	\cline{1-7}\\
	{}	&	\multicolumn{6}{c}{Blueshifted Components~(mJy)}&{}\\
	\cline{2-7}
	VLA	&	$-1.75(0.18)$	&	$-2.34(0.27)$	&	$1.54(0.45)$	&	$-1.60(0.20)$	&	$-1.23(0.15)$	&	$-2.07(0.24)$	&	\\
	GBT\footnotemark[$\ast$]	&	$-3.40(0.64)$	&	$-4.26(0.59)$	&	...	&	$-3.46(0.69)$	&	$-2.45(0.69)$	&	$-1.33(0.96)$	&	\\
	$|S_{\rm GBT}$/$S_{\rm VLA}|$ &	$1.95(0.42)$	&	$1.82(0.33)$	&	...	&	$2.16(0.51)$	&	$1.99(0.61)$	&	$0.64(0.47)$	&	\\
   \hline\\
   \multicolumn{3}{@{}l@{}}{\hbox to 0pt{\parbox{85mm}{\footnotesize
 \footnotemark[$\ast$] GBT data are from \citet{mangum2013}.
     }\hss}}
   \end{tabular}
 \end{center}
 \end{table}

\begin{table}
 \begin{center}
\caption{The emission lines close to NH$_3(3,3)$} 
\label{tab:nh3}
  \begin{tabular}{ccc}
   \hline
   \multicolumn{1}{c}{Transitions}		&	{Rest frequency (GHz)}	&	{$\Delta V$ (km~s$^{-1}$)\footnotemark[$\ast$]}\\
\hline
H(81)$\beta$	&	23.86086	&	117	\\
NH$_3(3,3)$	&	23.87013	&	---	\\
He(81)$\beta$	&	23.87059	&	$-6$	\\
\cline{1-3}
{}	&	{}	&	{Emission velocity range (km~s$^{-1}$)\footnotemark[$\dagger$]}\\
\cline{1-3}
NH$_3(3,3)$	&	23.87013	&	835--1045	\\
H(80)$\beta$	&	24.75574	&	$\simeq718$--928\\
He(80)$\beta$	&	24.76583	&	$\simeq841$--1051\\
   \hline\\
   \multicolumn{3}{@{}l@{}}{\hbox to 0pt{\parbox{100mm}{\footnotesize
\footnotemark[$\ast$] Velocity difference relative to the rest frequency of NH$_3(3,3)$.
      \par\noindent
\footnotemark[$\dagger$] Estimated velocity range in H(80)$\beta$ and He(80)$\beta$ in case that the emission features in figure~\ref{fig:nh3_33} are caused by H(81)$\beta$ and He(81)$\beta$, respectively.
     }\hss}}
   \end{tabular}
 \end{center}
 \end{table}

\begin{table}
 \begin{center}
\caption{OH parameters} 
\label{tab:fit_oh}
  \begin{tabular}{cccccccc}
   \hline
   \multicolumn{1}{c}{Transitions}		&	{$S$\footnotemark[$\ast$]}	&	{$V_{\rm LSR}$\footnotemark[$\ast$] }	&	{$\Delta v_{1/2}$}	&	{$\tau$\footnotemark[$\ast$]}	&	{$\int\tau dv$}	&		&	{$S_{\rm cont}$(rms noise)}	\\
{}	&	{(mJy)}	&	{(km~s$^{-1}$)}	&	{(km~s$^{-1}$)}	&		&	{(km~s$^{-1}$)}	&		&	{ (mJy)}	\\
\hline
{OH $^2\Pi_{2/3} $$J=9/2$}	&	\multicolumn{5}{c}{Systemic Components}	&	{}	&	{}	\\
\cline{2-6} 
$F=4-4$	&	$-2.23\pm0.16$		&	$1085\pm4$	&	$98\pm8$		&	$0.046\pm0.003$	&	$4.75\pm0.53$	&	&	49.99 (0.28)	\\
$F=5-5$	&	$-1.25\pm0.10$		&				&	$193\pm19$	&	$0.025\pm0.002$	&	$5.21\pm0.67$	&	&				\\\\
{}	&	\multicolumn{5}{c}{Refshifted Components}	&	{}	&	{}	\\
\cline{2-6} 
$F=4-4$	&	$-2.00\pm0.14$		&	$1302\pm2$	&	$62\pm6$		&	$0.041\pm0.003$	&	$2.68\pm0.34$	&	&				\\
$F=5-5$	&	$-3.51\pm0.14$		&				&	$107\pm7$	&	$0.073\pm0.003$	&	$8.31\pm0.64$	&	&				\\
   \hline\\
   \multicolumn{3}{@{}l@{}}{\hbox to 0pt{\parbox{100mm}{\footnotesize
\footnotemark[$\ast$]{Values at the absorption peak after Gaussian fitting.}
     }\hss}}
   \end{tabular}
 \end{center}
 \end{table}

\section{Discussion}
\subsection{The possibility of NH$_3(3,3)$ maser}\label{sec:nh3_maser}
It has been reported that distributions of NH$_3 (3,3)$ masers and H$_2$O masers in the massive star forming  regions are associated each other, 
because the NH$_3 (3,3)$ masers are formed by shocks due to the interaction of molecular outflows with the ambient gas in, e.g., NGC 6334 \citep{kraemer1995}, DR21 \citep{zhang1995}.
VLBI observations of H$_2$O maser  toward the center of {NGC 3079} also strongly support the possibility of the NH$_3$ maser.
\citet{yamauchi2004} found exceptionally strong blueshifted H$_2$O maser features at the nearly same velocities of $V_{\rm LSR}\approx900$--1050~km~s$^{-1}$ as our blueshifted components, showing peculiar motion (LSM1 and LSM2 in their paper) that does not follow the rotation of the nuclear maser disk. 
The velocity and  spatial distributions of the maser spots suggested two expanding partial shells
in the massive and geometrically thick maser disk and located at about 7~mas (0.7~pc; LSM2) and 13~mas (1.2~pc; LSM1) from the nucleus, 
being probably formed by local shock related to star formation \citep{yamauchi2004}.
These results strongly suggest that the emission features of NH$_3(3,3)$  are caused by the NH$_3(3,3)$ maser associated with shocks by winds from newly formed massive stars or supernova explosions in extremely dense clouds in the nuclear region of {NGC 3079}, although the distribution of the NH$_3(3,3)$ maser is unresolved with the angular resolution of $\sim0\farcs1$ [see figure~\ref{fig:map2}~(left)]. 
The blueshifted absorption features of NH$_3(1,1)$, $(2,2)$ and $(4.4)$--$(6,6)$ at the same velocities could be also caused by the dense molecular gas in the nuclear H$_2$O maser disk.

Absorption features of para-NH$_3$ have been found in IC~860, NGC~253, NGC~660, NGC~3079 and Arp~220 \citep{mangum2013, ott2005, takano2005, ott2011}, 
while maser emission of NH$_3(3,3)$ has been detected only in NGC~253 and NGC~3079 \citep{ott2005, mangum2013}.
\citet{ott2005} speculated that the maser emission in NGC~253 was associated with the star forming region. 
Measurements of the brightness temperature and distribution of the emission with VLBI could determine if the emission of NH$_3(3,3)$ is masers associated with star formation.

\subsection{Rotational Temperatures of NH$_3$}\label{sec:trot}
In the case of an absorption line, the optical depth can be derived from the brightness temperature of the  line ($T_{\rm L}$) relative to the measured  brightness temperature of the background continuum  $T'_{\rm C} (=f_{\rm C} T_{\rm C}$) as below (e.g., \cite{huttemeister1993,ott2011}),  
\begin{eqnarray}
\label{eq:tau}
\tau = -\ln\left( 1-\frac{|T_{\rm L}|}{T'_{\rm C}}\right), 
\end{eqnarray}
where $T_{\rm C}$ is the continuum brightness temperature and $f_{\rm C}$ the beam filling factor of the continuum source whose individual emitting regions are small as shown with VLBI (e.g., \cite{trotter1998, yamauchi2004}) and thus which is assumed to be fully covered by the molecular gas.
The resultant peak optical depths of NH$_3$ and OH are given in table~\ref{tab:fit_nh3} and \ref{tab:fit_oh}, respectively.
Using the optical depth, the NH$_3$ column density divided by the excitation temperature ($T_{\rm ex}$) for the inversion doublet can be expressed by 
\begin{eqnarray}
\frac{N(J,K)}{T_{\rm ex}}=1.65\times10^{14} \frac{J(J+1)}{K^2 \nu}\tau \Delta v_{1/2}~[{\rm cm}^{-2} {\rm K}^{-1}] , 
\label{eq:column}
\end{eqnarray}
as in \citet{mauersberger1986b}, where $\nu$ is the transition frequency in GHz  
and $\Delta v_{1/2}$ the line width (FWHM) in km~s$^{-1}$. 
The NH$_3$ populations are described by two temperatures, one is the excitation temperature $T_{\rm ex}$ which characterizes the population across a $(J,K)$ inversion doublet and the other the rotational temperature  $T_{\rm rot}$ which characterizes the populations of energy levels with different $(J,K)$. 
Since the populations of metastable inversion levels are determined by collisions, 
there is a direct relation between $T_{\rm rot}$ and the kinetic temperature $T_{\rm k}$. 
The column density divided by $T_{\rm ex}$ 
as a function of energy is described by the Boltzmann law with a specific $T_{{\rm rot} (J',K'; J, K)}$, 
\begin{eqnarray}
\frac{N(J',K')/T'_{\rm ex}}{N(J,K)/T_{\rm ex}}&=&
\frac{g_{\rm op'}} {g_{\rm op}} \frac{(2J'+1)}{(2J+1)}\exp\left(-\frac{\Delta E}{T_{{\rm rot}(J',K'; J, K)}}\right),  
\end{eqnarray}
as in  \citet{ott2011}, where $\Delta E$ is the energy difference between the rotational states $(J',K')$ and $(J, K)$ in K, and 
$g_{\rm op}$ the statistical weight factor 
[$g_{\rm op}=1$ for para-NH$_3$ ($K\neq3n$) and $K=0$, and $g_{\rm op}=2$ for ortho-NH$_3$ ($K=3n$)]. 

The rotational temperature can be derived from the rotational diagram (figure~\ref{fig:Trot}), which is the logarithmic plot of the normalized column densities, $\log[N(J,K)/(T_{\rm ex} (2J+1) g_{\rm op})]$ as a function of energy levels corresponding to the transitions. 
Here we treat ortho- and para-NH$_3$ separately, because 
the transitions between ortho- and para-NH$_3$ are not allowed and hence the two behave independently.
The rotational temperatures for the systemic component derived from two transitions of  NH$_3 (1,1)$--$(2,2)$ and $(4,4)$--$(5,5)$ are $T_{\rm rot~(11,22)}= 64\pm5$~K and $T_{\rm rot~(44,55)}= 164\pm14$~K. 
This indicates that the systemic components consist of at least two temperature components, like  the center of the Galaxy (e.g., \cite{huttemeister1995, mills2013}) and the other galaxies, 
(e.g., {NGC~1068}: \cite{ao2011}). 
The presence of two temperature components in the central region of {NGC~3079}, i.e., the hot molecular gas $(T_{\rm k}= 150$~K, $n_{\rm H_2} = 10^3$~cm$^{-3})$ and the cold dense gas $(T_{\rm k} = 20$~K, $n_{\rm H_2} = 10^4$~cm$^{-3})$,  
  was also shown by the radiative transfer model
using multi-transition of $^{12}$CO and $^{13}$CO lines with single dishes ($\theta\sim23\arcsec$ in CO(1--0); \cite{israel2009}).
Moreover in figure~\ref{fig:Trot}, the larger column density of NH$_3(6,6)$, whose energy level is 408~K, than  the value extrapolated from the lower transition lines indicates the existence of the hotter gas. 

We also derived the mean rotational temperature of $T_{\rm rot}=120\pm12$~K for the systemic features,  using para-NH$_3$ lines, $(1,1)$, $(2,2)$, $(4,4)$ and $(5,5)$, for comparing with other galaxies. 
The  derived rotational temperature is consistent with the previous results of {NGC~3079} [$T_{\rm rot}>106$~K, derived from the same para-NH$_3$ lines in \citet{mangum2013}], 
and is higher than temperatures in  other galaxies already reported (e.g., $T_{\rm rot}=44^{+6}_{-4}$~K in {NGC~1068}; \cite{ao2011}).
The rotational temperature of the blueshifted component was evaluated to be $T_{\rm rot}=157\pm19$~K from the para-NH$_3$ lines. 

The kinetic temperature ($T_{\rm k}$) is, in general, higher than the rotational temperature   \citep{walmsley1983}. 
The relation between $T_{\rm rot}$ and $T_{\rm k}$, which was derived from radiative transfer large velocity gradient (LVG) models, was shown by \citet{ott2011} for ammonia rotational transitions.
We roughly estimate the mean kinetic temperature of $T_{\rm k}=270\pm60$~K for the systemic components using  the mean rotational temperature, 
$T_{\rm rot}=120$~K, and 
the relation between $T_{\rm k}$ and $T_{\rm rot~(22,44)}$ (figure~5 in \cite{ott2011}).
For the blueshifted components,  
we can derive 
the lower limit of the kinetic temperature, $T_{\rm k}>500$~K.

\subsection{Rotational Temperature of OH}\label{sec:trot_oh}
We also estimated the rotational temperature of the systemic component of OH from the ratio of the normalized column density between  $^2\Pi_{3/2}$~$J=3/2$ and $J=9/2$, assuming that the systemic components of OH~$^2\Pi_{3/2}$~$J=3/2$ in the 1667- and 1665-MHz transitions \citep{hagiwara2004} are emitted from the same region as that of OH~$^2\Pi_{3/2}$~$J=9/2$. 
The normalized OH column density is expressed as \citep{baudry1981},
\begin{eqnarray}
\frac{N_l}{T_{\rm ex}}=2.07\times10^3\nu^2\frac{g_l}{g_u}A_{ul}^{-1}\tau\Delta v_{1/2}~[{\rm cm}^{-2}~{\rm K}^{-1}], 
\end{eqnarray}
where $N_l$ is the column density of one ($F=4$--4 or 5--5) of the two hyperfine levels at the lower state of the $\Lambda$-doublet, $\nu$  the frequency of the transition in GHz, $A_{ul}$ the Einstein A-coefficient \citep{destombes1977}, and $g_l$ and $g_u$ are the statistical weights $(2F+1)$ of the lower and upper levels, respectively.
Since the mean total column density of $J=9/2$ is given by 
\begin{eqnarray}
\frac{N_{9/2}}{T_{\rm ex}}=7.46\times10^{13} \tau_{4-4}\Delta v_{1/2}+7.42\times10^{13} \tau_{5-5}\Delta v_{1/2}~[{\rm cm^{-2}}~{\rm K^{-1}}], 
\end{eqnarray}
as in \citet{baudry1995}, we obtain $N_{9/2}/T_{\rm ex}=6.98\times 10^{14}$~cm$^{-2}$~K$^{-1}$ for the systemic component, using the parameters of $\tau$ and $\Delta v_{1/2}$ for $F=4$--4 and 5--5 in table~\ref{tab:fit_oh}.
The rotational temperature of $T_{\rm rot, OH}\sim175$~K between $J=9/2$ and $J=3/2$  was derived from the  equation,  
\begin{eqnarray}
\left(N_{9/2}/T_{\rm ex}\right)/\left(N_{3/2}/T'_{\rm ex}\right)=(10/4)\exp(-\Delta E/T_{\rm rot})~,
\end{eqnarray}
where $N_{3/2}/T'_{\rm ex} = 5.3\times 10^{15}$~cm$^{-2}$~K$^{-1}$ \citep{hagiwara2004}. 
The rotational temperature of OH has large uncertainty because the velocities of the systemic components between OH~$^2\Pi_{3/2}$~$J=3/2$ and OH~$^2\Pi_{3/2}$~$J=9/2$ lines are not exactly same [compare our figure~\ref{fig:oh} with figure~1 of \citet{hagiwara2004}].
It is expected that 
the rotational temperature of the redshifted component could be higher than that of the systemic, 
because the OH~$^2\Pi_{3/2}$~$J=3/2$ in ground state was not detected but the $^2\Pi_{3/2}$~$J=9/2$ whose energy level is 511~K.

\subsection{Distribution of molecular gas in the nuclear region}\label{subsec:distribution}
Figure~\ref{fig:peak} shows the distribution of the peak positions of NH$_3$ and OH lines for the systemic and blueshifted components in the coordinate system relative to the position of  R.A.~(J2000.0)$=10^{\rm h} 01^{\rm m} 57\fs8034$, Decl.~(J2000.0)$=55$\arcdeg40\arcmin47\farcs243 determined from observations with VLBI \citep{petrov2011}.
The statistic position errors were evaluated from $\Delta\theta=0.5~\theta_{\rm beam}/{\it SNR}$ (e.g., \cite{moran1999}), where $\theta_{\rm beam}$ is the synthesized beam size (table~\ref{tab:obs1}) and $SNR$ the signal-to-noise ratio of the peak absorption. 
The circles in figure~\ref{fig:peak} show the positions of the peak intensities of the NH$_3(1,1)$, $(2,2)$, $(4,4)$, $(5,5)$ and $(6,6)$ absorption integrated with 2 and 3 channels, corresponding to $V_{\rm LSR}=1105$--1135~km~s$^{-1}$ and 1000--1045~km~s$^{-1}$,  for the systemic and blueshifted components, respectively.
The peak positions of the NH$_3(3,3)$ lines integrated with $V_{\rm LSR}=1120$--1165~km~s$^{-1}$ for the systemic components (absorption) and 970--1045~km~s$^{-1}$ for the blueshited (emission) are also plotted.

In figure~\ref{fig:peak}, the systemic and blueshifted features, with the exception of blueshifted NH$_3(1,1)$ and $(3,3)$ are located at the north-west and south-east relative to the reference center, respectively. 
A least-squares fitting of the systemic and blueshifted components gives ${\it P.A.}\approx-53\arcdeg$, 
in agreement with the position angle of the nuclear continuum jet measured with VLBI (${\it P.A.}=-52\arcdeg$, \cite{trotter1998, yamauchi2004}), 
suggesting that ammonia is absorbed against the continuum emission of the nuclear jet.
The velocity structure of NH$_3$ absorption is inconsistent with the kpc-scale galactic disk and the pc-scale nuclear water maser disk with ${\it P.A.}=-8\arcdeg$ which rotates with the northern side approaching and the southern receding \citep{yamauchi2004}, but is nearly consistent with that of OH absorption at 1665-MHz \citep{hagiwara2004}.
The projected separation between the averaged positions of the systemic and blueshifted components in figure~\ref{fig:peak}, 8.6~mas, is also consistent with the position offsets of $\sim7$--8~mas between two peaks of the systemic and blueshifted components of the OH absorption \citep{hagiwara2004}.
The blueshifted absorption features of NH$_3$ and OH could be explained with an association with molecular outflows possibly caused by the jet, 
as suggested by \citet{hagiwara2004}.

On the other hand, the relative position of the blueshifted NH$_3(3,3)$ emission is located at $\sim10$~mas north of the nuclear jet, in agreement with the positions of the H$_2$O masers (LSM2) formed by local shocks related to star formation in the nuclear maser disk \citep{yamauchi2004}.
The position of the blueshifted NH$_3(1,1)$ close to the NH$_3(3,3)$ could be explained by the enhancement of NH$_3$ due to the local shock (e.g., \cite{draine1993, flower1995}). 

Figure~\ref{fig:peak} also shows the peak positions of the systemic and redshifted components of the OH~$^2\Pi_{3/2}$~$J=9/2$~$F=4$--4 and $F=5$--5 lines, where 
the velocity ranges of $V_{\rm LSR}=1075$--1105~km~s$^{-1}$ and 1285--1315~km~s$^{-1}$ are adopted for  the systemic and redshifted components of  the $F=4$--4 line, and  $V_{\rm LSR}=970$--1000~km~s$^{-1}$ and 1180--1210~km~s$^{-1}$ for  the systemic and  redshifted of  the $F=5-5$ line. 
The peak positions of the systemic components of the $F=4$--4 and $F=5$--5 are located  close to the positions of the systemic NH$_3$ lines.
The redshifted component of the OH~$F=5$--5 is located at the northwestern side with respect to the systemic components of the OH~$F=4$--4, 5--5, and NH$_3$ lines aligned along the nuclear jet.
The redshifted component of the OH~$F=4$--4 line is however at the further northern side near the shock gas region of H$_2$O masers \citep{yamauchi2004} and the blueshifted  component of NH$_3(3,3)$.
To make the distributions of these molecular gas clear, observations with higher angular resolution, e.g, with VLBI would be needed.

\subsection{Abundance of Ammonia}\label{subsec:column}
\begin{figure}
 \begin{center}
  \includegraphics[width=80mm]{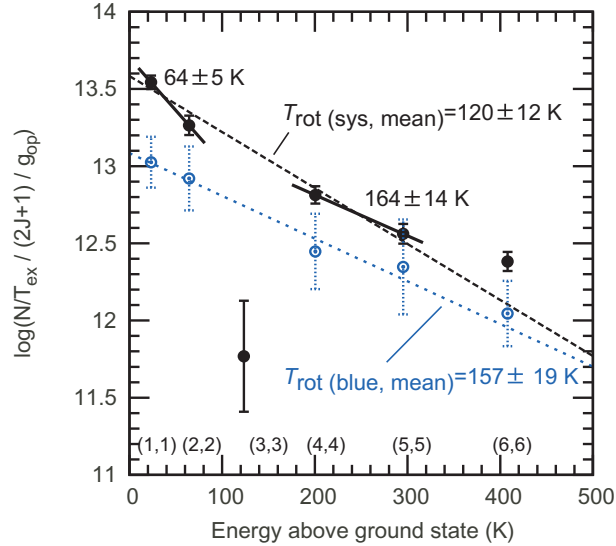} 
 \end{center}
\caption{Rotational diagram of the ammonia measurements. 
The ordinate exhibits the logarithm of normalized $N/T_{\rm ex}$. 
The systemic and  blueshifted components 
are marked by filled  and open circles, respectively. 
The thick and thin dotted lines indicate the results of the least-square fitting of the systemic and blueshifted components, respectively, for $(J,K)=(1,1), (2,2), (4,4)$ and $(5,5)$.
The mean rotational temperatures for the systemic and blueshifted features are shown.
In addition, the rotational temperatures derived from two transitions of NH$_3(1,1)$--$(2,2)$ and $(4,4)$--$(5,5)$ for the systemic component are shown.
}
 \label{fig:Trot}
 \end{figure}

\begin{figure}
 \begin{center}
  \includegraphics[width=80mm]{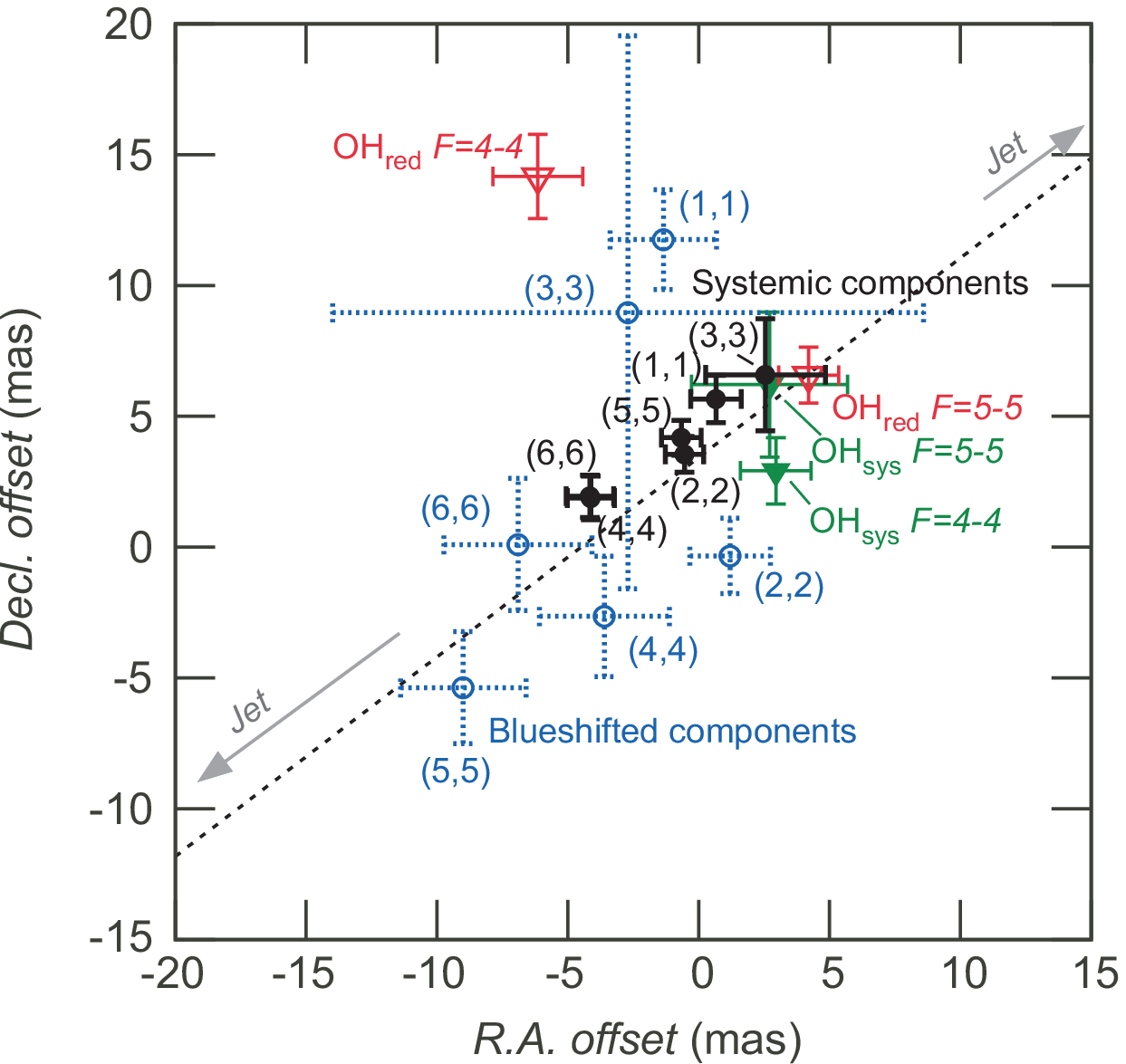} 
 \end{center}
\caption{Distribution of NH$_3$ lines in the nuclear region of {NGC~3079}. 
The origin of the coordinates is R.A.~(J2000.0)$=10^{\rm h} 01^{\rm m} 57\fs8034$, Decl.~(J2000.0)$=55$\arcdeg40\arcmin47\farcs243 \citep{petrov2011}.
Filled and open circles indicate the peak positions of NH$_3(1,1)$--$(6,6)$ for the systemic and blueshifted components, respectively. 
A dashed line with a position angle of ${\it P.A.}=-53\arcdeg$ is the result of a least-squares fitting of the positions with the exception of blueshifted NH$_3(1,1)$ and $(3,3)$. 
Filled triangles show the peak positions of the systemic components of OH~$^2\Pi_{3/2}$~$J=9/2$~$F=4$--4 and $F=5$--5, 
and open triangles those of the redshifted. }
 \label{fig:peak}
 \end{figure}

In order to derive column densities of NH$_3$ from the absorption lines, the excitation temperature $T_{\rm ex}$ is required, as shown  in equation~(\ref{eq:column}).
Although radiative transfer models are needed to determine $T_{\rm ex}$ (e.g., \cite{walmsley1983}), 
we estimated the column densities for the systemic and blueshifted components, 
assuming  $T_{\rm ex}\approx T_{\rm rot} (< T'_{\rm C} \sim 1.7 \times 10^4$~K; $T'_{\rm C}$ $=$ the brightness temperature of the background continuum source).
The ortho-NH$_3 (0,0)$ does not exhibit an inversion line and $(3,3)$ is degenerated by the contamination of the maser emission (subsection~\ref{subsec:ammonia}), 
and hence we extrapolate the column densities of NH$_3(0,0)$ and $(3,3)$ from the rotational temperature for para-NH$_3$ and $N/{T_{\rm ex}}$ of NH$_3(6,6)$ in figure~\ref{fig:Trot}.
The derived column densities of the systemic and blueshifted components are summarized in table~\ref{tab:column}. 

The fractional abundances of $\sum^6_{J=0} N_J~({\rm NH_3})$ 
relative to the column densities of H$_2$ for the systemic and blueshifted components are [NH$_3$]/[H$_2]=1.3\times10^{-7}$ and $6.5\times10^{-8}$, respectively,  
where  the column density of H$_2$ ($=6.8 \times 10^{23}$~cm$^{-2}$) is estimated from the CO integrated intensity, $I_{\rm CO}\equiv \int T_{b}~dv$, with the angular resolution of $1\farcs9$ \citep{koda2002} and the conversion factor of $X\equiv N({\rm H}_2)/I_{\rm CO}=1.8\times 10^{20}$~cm$^{-2}$~(K~km~s$^{-1}$)$^{-1}$ 
\citep{dame2001}.
It is difficult to decide whether the difference of the abundance between the systemic and blueshifted components by a factor of two is real, because the column densities of NH$_3(3,3)$ estimated only from that of NH$_3(6,6)$ and $T_{\rm rot}$ of para-NH$_3$ occupy 46~\% and 42~\% of the total column densities of the systemic and blueshifted component, respectively, maybe causing the uncertainty of nearly a factor of two for the difference.

In the central regions ($r\leq0.17$--1.1~kpc) of other galaxies, the fractional abundance of ammonia was (1.3--$2.9)\times10^{-8}$ for NGC~253, IC~342, Maffei~2 and NGC~1068, $4.5\times10^{-9}$ for M51, and $5\times10^{-10}$ for M82 (see \cite{takano2013} and references therein).
The ammonia abundance, at least of the systemic component $(1.3\times10^{-7})$, in the nuclear region of NGC~3079  is higher than that of the other galaxies, even if the uncertainty of the density of NH$_3(3,3)$ is considered.
To know the reason of the high abundance in NGC~3079, 
it would be important to measure the ammonia abundance in the nuclear regions of other galaxies, especially active galactic nuclei having jet, from observations with high resolutions of $r\leq10$~pc.

\begin{table}
 \begin{center}
	\caption{Ammonia rotational temperatures and column densities} 
	\label{tab:column}
	\begin{tabular}{ccc}
	\hline
   \multicolumn{1}{c}	{}			&	{Systemic component}	&	{Blueshifted component} 	\\
	{$T_{\rm rot}$}	&	$120\pm12$~K	&	$157\pm19$~K	\\
	\cline{1-3}\\
	{}			&	\multicolumn{2}{c}{Column density\footnotemark[$\ast$]~[$\times10^{15}$ cm$^{-2}$]} 	\\
	{NH$_3(0,0)$}	&	$(4.6\pm0.5)$\footnotemark[$\dagger$]	&	$(1.9\pm0.2)$\footnotemark[$\dagger$]	\\
	{NH$_3(1,1)$}	&	$12.5\pm0.5$	&	$5.0\pm0.8$	\\
	{NH$_3(2,2)$}	&	$11.0\pm0.7$	&	$6.5\pm1.4$	\\
	{NH$_3(3,3)$}	&	$(41.1\pm4.1)$\footnotemark[$\dagger$]	&	$(19.0\pm2.3)$\footnotemark[$\dagger$]	\\
	{NH$_3(4,4)$}	&	$7.0\pm0.4$	&	$4.0\pm1.0$	\\
	{NH$_3(5,5)$}	&	$4.8\pm0.3$	&	$3.8\pm1.2$	\\
	{NH$_3(6,6)$}	&	$7.5\pm0.5$	&	$4.5\pm1.0$	\\
	\cline{1-3}\\
	{$\displaystyle \sum^6_{J=0} N$}	&	$88.5\pm7.0$		&	$44.7\pm7.8$ \\
	   \hline\\
	\multicolumn{3}{@{}l@{}}{\hbox to 0pt{\parbox{100mm}{\footnotesize
	\footnotemark[$\ast$]{Assumed $T_{\rm ex} \approx T_{\rm rot}$.}
       \par\noindent
	\footnotemark[$\dagger$]{Brackets are values extrapolated from $T_{\rm rot}$ for para-ammonia and $N/T_{\rm ex}$ of NH$_3(6,6)$.}
     }\hss}}
   \end{tabular}
 \end{center}
 \end{table}

\section{Conclusions}
We reported the results of ammonia observations toward the center of {NGC~3079} with 
the Tsukuba 32-m telescope and VLA.
The main conclusions are summarized as follows:
\begin{enumerate}

\item 
We detected ammonia $(J,K)=(1, 1)$ and $(2, 2)$  inversion lines in absorption with the Tsukuba 32-m, 
and  $(J,K)=(1, 1)$ through $(6, 6)$ lines with the VLA although the profile of the NH$_3(3,3)$ line was anomalous.

\item 
We detected an unresolved continuum source with the resolution of $\lesssim0\farcs09\times0\farcs08$.
On the other hand, the flux density of $\sim50$~mJy is smaller than the  flux obtained by observations with the GBT, 
and  larger than that with VLBI. 
These results indicate that the continuum source is extended  more than $r\sim1\arcsec$. 

\item 
All ammonia absorption lines have two distinct velocity components: 
one is at the systemic velocity and the other is blueshifted, 
and  align along the nuclear jet.
For the systemic components, the ratio of the flux densities with GBT to VLA decreases with the excitation level, indicating that the relatively low temperature gas is more extended than the high temperature gas.

\item 
At the blueshifted velocity, the NH$_3(3, 3)$ line shows emission feature. 
The blueshifted  feature is consistent with the distribution and velocity of the H$_2$O maser formed by local shock related to star formation.
We regarded NH$_3(3, 3)$ maser as the most likely candidate of the emission, and  
suggested that the  maser was associated with shocks by strong winds from newly formed massive stars or
supernova explosions in  dense clouds in the nuclear region of NGC 3079.

\item 
Using para-NH$_3(1, 1)$, (2, 2), (4, 4) and (5, 5) lines with the VLA, 
we derived the rotational temperature 
$T_{\rm rot} = 120 \pm 12$~K and $157 \pm 19$~K for the systemic and blueshifted components, respectively. 
The temperatures correspond to the kinetic temperatures of $T_{\rm k} = 270\pm60$~K and $> 500$~K for the systemic and blueshifted components, respectively.

\item
Combined column densities of the measured lines plus the extrapolated $(0, 0)$ column 
become $(8.85\pm0.70) \times 10^{16}$~cm$^{-2}$ and $(4.47 \pm 0.78) \times 10^{16}$~cm$^{-2}$ for the systemic and blueshifted components, respectively, assuming $T_{\rm ex}\approx T_{\rm rot}$.
The fractional abundances of $\sum^6_{J=0} N_J~({\rm NH_3})$ 
relative to the column densities of H$_2$ for the systemic and blueshifted components are [NH$_3$]/[H$_2]=1.3\times10^{-7}$ and $6.5\times10^{-8}$, respectively.

\item 
We found an absorption which is likely to be caused by the $F=4$--4 and $F=5$--5 doublet of OH~$^2\Pi_{3/2}$~$J=9/2$. 
Given that each doublet line has same velocity components, the wide absorption feature can be fitted by two velocity components. 
These velocity components are different from the velocities at the ground state of OH~$^2\Pi_{3/2}$~$J=3/2$ and NH$_3$, 
not showing the blueshifted components but the redshift. 
We estimated the rotational temperature 
of $T_{\rm rot, OH} \gtrsim 175$~K from OH~$^2\Pi_{3/2}$ data. 
The highly excited OH line can trace hot gas 
associated with the interaction of the fast nuclear outflow with dense and dusty molecular material around the nucleus.

\end{enumerate}

\bigskip
We thank the members of the observational astrophysics group of the University of Tsukuba for participating in the construction of the K-band observing system of the Tsukuba 32-m telescope.
We are grateful to the VLBI group of the Geospatial Information Authority of Japan (GSI) geodetic department for licensing to use the 32-m telescope.
The observations with the 32-m telescope have been made under the agreement on the collaboration between the University of Tsukuba and the GSI.
This work was supported in part by JSPS (Japan Society for the Promotion of Science) KAKENHI Grant Numbers 173400052, 17654042, 20244011 and 26247019.
The National Radio Astronomy Observatory is a facility of the National Science Foundation operated under cooperative agreement by Associated Universities, Inc.
This research has made use of the NASA/IPAC Extragalactic Database (NED) which is operated by the Jet Propulsion Laboratory, California Institute of Technology, under contract with the National Aeronautics and Space Administration.
Some of the data presented in this paper were obtained from the Mikulski Archive for Space Telescopes (MAST). STScI is operated by the Association of Universities for Research in Astronomy, Inc., under NASA contract NAS5-26555. Support for MAST for non-HST data is provided by the NASA Office of Space Science via grant NNX13AC07G and by other grants and contracts.



\begin{thebibliography}{}
\bibitem[Ao et al.(2011)] {ao2011} Ao,~Y., Henkel,~C., Braatz,~J.~A., Wei{\ss},~A., Menten,~K.~M. \& M{\"u}hle,~S. 2011, \aap, 529, 154

\bibitem[Baan \& Irwin(1995)] {baan1995} Baan,~W.~A. \& Irwin,~J.~A. 1995, \apj, 446, 602
\bibitem[Baudry et al.(1981)] {baudry1981} Baudry,~A., Walmsley,~C.~M., Winnberg,~A. \& Wilson,~T.~L. 1981, \aap, 102, 287
\bibitem[Baudry \& Menten(1995)] {baudry1995} Baudry,~A. \& Menten,~K.~M. \aap, 298, 905
\bibitem[Braine et al.(1997)] {braine1997} Braine,~J., Guelin,~M., Dumke,~M., Brouillet,~N., Herpin,~F. \& Wielebinski,~R. 1997, \aap, 326, 963
\bibitem[Brunthaler et al.(2008)] {brunthaler2008} {Brunthaler},~A., {Castangia},~P., {Tarchi},~A., {Henkel},~C., {Reid},~M.~J., {Falcke},~H. \& {Menten},~K.~M. 2009, \aap, 497, 103
	
\bibitem[Cecil et al.(2001)] {cecil2001}  Cecil,~G., Bland-Hawthorn,~J., Veilleux,~S. \& Filippenko,~A.~V. 2001, \apj, 555, 338

\bibitem[Dame et al.(2001)] {dame2001} Dame,~T.~M., Hartmann,~D., \& Thaddeus,~P. 2001, \apj, 547, 792
\bibitem[Danby et al.(1988)] {danby1988} Danby,~G., Flower,~D.~R., Valiron,~P., Schilke,~P. \& Walmsley,~C.~M. 1988, \mnras, 235, 229
\bibitem[de Pater et al.(2005)] {depater2005} de Pater,~I., Gibbard,~S.~G.,  Chiang,~E.,  Hammel,~H.~B., Macintosh,~B., Marchis,~F., Martin,~S.~C.,  Roe,~H.~G. \& Showalter,~M. 2005, Icarus, 174, 263
\bibitem[Destombes et al.(1977)] {destombes1977} Destombes,~J.~L., Marliere,~C., Baudry,~A., \& Brillet,~J. 1977, \aap, 60, 55
\bibitem[de Vaucouleurs et al.(1991)] {devaucouleurs1991} de Vaucouleurs,~G., de Vaucouleurs,~A., Corwin,~Jr.,~H.~G., Buta,~R.~J., Paturel,~G. \& Fouqu{\'e},~P. 1991, Third Reference Catalogue of Bright Galaxies (Springer, New York) 
\bibitem[Draine \& McKee(1993)] {draine1993} Draine,~B.~T. \& McKee,~C.~F. 1993, \araa, 31, 373
\bibitem[Duric et al.(1988)] {duric1988} Duric,~N., \& Seaquist,~E., 1988, \apj, 326, 574

\bibitem[Flower et al.(1995)] {flower1995} Flower,~D.~R., Pineau des Forets,~G. \& Walmsley,~C.~M. 1995, \aap, 294, 815
\bibitem[Ford et al.(1986)] {ford1986} Ford,~H.~C., Dahari,~O., Jacoby,~G.~H., Crane,~P.~C. \& Ciardullo,~R, 1986, \apj, 311, L7

\bibitem[Guilloteau et al.(1984)] {guilloteau1984} Guilloteau,~S., Baudry,~A., Walmsley,~C.~M., Wilson,~T.~L. \& Winnberg,~A. 1984, \aap, 131, 45

\bibitem[Hagiwara et al.(2004)] {hagiwara2004} Hagiwara,~Y., Kl{\"o}ckner,~H.-R. \& Baan,~W., 2004, \mnras, 353, 1055
\bibitem[Hawarden et al.(1995)] {hawarden1995} Hawarden,~T.~G., Israel,~F.~P., Geballe,~T.R. \& Wade, R., 1998, \mnras, 276, 1197
\bibitem[Heckman(1980)] {heckman1980} Heckman,~T.~M. 1980, \aap, 87, 152
\bibitem[Heckman et al.(1990)] {heckman1990} Heckman,~T.~M., Armus,~L. \& Miley,~G.~K. 1990, \apjs, 74, 833

\bibitem[Henkel et al.(2000)] {henkel2000} Henkel,~C., Mauersberger,~R., Peck,~A.~B., Falcke,~H. \& Hagiwara,~Y. 2000 \aap, 361, 45

\bibitem[H\"{u}ttemeister et al.(1993)] {huttemeister1993} H\"{u}ttemeister,~S., Wilson,~T.~L., Henkel,~C.  \& Mauersberger,~R.  1993, \aap, 276, 445 
\bibitem[H\"{u}ttemeister et al.(1995)] {huttemeister1995} H\"{u}ttemeister,~S., Wilson,~T.~L., Mauersberger,~R., Lemme,~C., Dahmen,~G. \& Henkel,~C., 1995, \aap, 300, 636 

\bibitem[Irwin \& Seaquist(1991)] {irwin1991} Irwin,~J.~A. \& Seaquist,~E.~R. 1986, \apj, 371, 111
\bibitem[Irwin \& Sofue(1992)] {irwin1992} Irwin,~J.~A. \& Sofue,~Y. 1992, \apj, 396, 75
\bibitem[Israel et al.(1998)] {israel1998} Israel,~F.~P., van der Werf,~P.~P., Hawarden,~T.~G., \& Aspin,~C., 1998, \aap, 336, 433
\bibitem[Israel (2009)] {israel2009} Israel, F.~P. 2009, \aap, 493, 525

\bibitem[Koda et al.(2002)] {koda2002} Koda,~J., Sofue,~Y., Kohno,~K., Nakanishi,~H., Onodera,~S., Okumura,~S.~K. \& Irwin,~J.~A., 2002, \apj, 573, 105 
\bibitem[Kohno et al.(2001)] {kohno2001} Kohno,~K., Matsushita,~S., Vila-Vilar{\'o},~B., Okumura,~S.~K., Shibatsuka,~T., Okiura,~M., Ishizuki,~S. \& Kawabe,~R. 2001, ASPC, 249, 672
\bibitem[Kraemer \& Jackson(1995)] {kraemer1995} Kraemer,~K.~E. \& Jackson,~J.~M., 1995, \apj, 439L, 9

\bibitem[Mangum et al.(2013)] {mangum2013} Mangum,~J.~G., Darling, ~J., Henkel,~C., Menten,~K.~M., MacGregor,~M., Svoboda,~B.~E., Schinnerer,~E., 2013, \apj, 779, 33
\bibitem[Mauersberger et al.(1986)] {mauersberger1986b} Mauersberger,~R., Henkel,~C., Wilson,~T.~L. \& {Walmsley}, C.~M. 1986b, \aap, 162, 199  
\bibitem[McMullin et al.(2007)] {mcmullin2007} McMullin,~J.~P., Waters,~B., Schiebel,~D.,Young, W., \& Golap,~K., 2007, Astronomical Data Analysis Software and Systems XVI, 376, 127
\bibitem[Meaburn et al.(1998)] {meaburn1998} {Meaburn}, J., {Fernandez}, B.~R., {Holloway}, A.~J., {Pedlar}, A., {Mundell}, C.~G. and {Geballe}, T.~R., 1998, \mnras, 295, 45M
\bibitem[Middelberg et al.(2007)] {middelberg2007} Middelberg,~E., Agudo,~I., Roy,~A.~L. \& Krichbaum,~T.~P. 2007, \mnras, 377, 731
\bibitem[Mills \& Morris(2013)] {mills2013} Mills,~E.~A.~C. \& Morris,~M.~R., 2013, \apj, 772, 105
\bibitem[Moran et al.(1999)] {moran1999} {Moran}, J.~M., {Greenhill}, L.~J. and {Herrnstein}, J.~R. 1999, JApA, 20, 165

\bibitem[Nishiyama et al.(2001)] {nishiyama2001} Nishiyama,~K., Nakai,~N. \& Kuno,~N. 2001, \pasj, 63, 755 

\bibitem[Oka et al.(1971)] {oka1971}    Oka,~T., Shimizu,~F.~O., Shimizu,~T. \& Watson,~J.~K.~G. 1971, \apjl, 165, 15
\bibitem[Ott et al.(2005)] {ott2005} Ott,~J., Weiss,~A., Henkel,~C. \& Walter,~F., 2005, \apj, 629, 767
\bibitem[Ott et al.(2011)] {ott2011} Ott,~J., Henkel,~C., Braatz,~J.~A. \& Wei\ss,~A., 2011, \apj, 742, 95

\bibitem[Petrov \& Taylor(2011)] {petrov2011} Petrov,~L. \& Taylor,~G.~B., 2011, \aj, 142, 89

\bibitem[Sawada-Satoh et al.(2001)] {sawada2001} Sawada-Satoh,~S., Inoue,~M., Shibata,~K.~M., Kameno,~S.,  Nakai,~N. \& Migenes, V., 2001, in IAU Symp. 205, Galaxies and their Constituents at the Highest Angular Resolutions, ed. R. T. Schilizzi, 196
\bibitem[Sofue \& Irwin(1992)] {sofue1992} Sofue,~Y. \& Irwin,~J.~A. 1992, \pasj, 44, 353
\bibitem[Sofue et al.(2001)] {sofue2001} Sofue,~Y., Koda,~J., Kohno,~K.,  Okumura,~S.~K., Honma,~M., Kawamura,~A. \&  Irwin,~J.~A.  2001, \apj, 547, L115
\bibitem[Springob et al.(2009)] {springob2009} Springob,~C.~M., Masters,~K.~L., Haynes,~M.~P., Giovanelli,~R. \& Marinoni,~C. 2009, \apjs, 182, 474 

\bibitem[Takano et al.(2005)] {takano2005} Takano,~S, Nakanishi=K, Nakai,~N. \& Takano,~T. 2005, \pasj, 57, L29    
    
\bibitem[Takano et al.(2013)] {takano2013} Takano,~S, Takano,~T.,  Nakai,~N., Kawaguchi,~K. \& Schilke,~P. 2013, \aap, 552, 34 
\bibitem[Trotter et al.(1998)] {trotter1998} Trotter,~A.~S., Greenhill,~L.~J., Moran,~J.~M., Reid,~M.~J., Irwin,~J.~A. \& Lo,~K.-Y. 1998, \apj, 495, 740

\bibitem[Ulich \& Haas(1976)] {ulich1976} Ulich,~B.~L. \& Haas,~R.~W. 1976, \apjs, 30, 247

\bibitem[Veilleux et al.(1994)] {veilleux1994} Veilleux,~S., Cecil,~G., Bland-Hawthorn,~J., Tully,~R.~B., Filippenko,~A.~V.\& Sargent,~W.~L.~W. 1994, \apj, 433, 48

\bibitem[Walmsley \& Ungerechts(1983)] {walmsley1983} Walmsley,~C.~M. \& Ungerechts,~H. 198, \aap, 122, 16
\bibitem[Walmsley et al.(1986)] {walmsley1986} Walmsley,~C.~M., Baudry,~A., Guilloteau,~S., \& Winnberg,~A., 1986, \apj, 167, 151 

\bibitem[Wei{\ss} et al.(2001)] {weiss2011} Wei{\ss},~A., Neininger,~N., Henkel,~C., Stutzki,~J., \& Klein,~U. 2001 \apj, 554L, 143W
	
\bibitem[Winnberg et al.(1978)] {winnberg1978} Winnberg,~A., Walmsley,~C.~M. \& Churchwell,~E. 1978, \aap, 66, 431

\bibitem[Yamagishi et al.(2010)] {yamagishi2010} Yamagishi,~M., Kaneda,~H., Ishihara,~D., Komugi,~S., Suzuki,~T. \& Onaka,~T. 2010, \pasj, 62, 1085   
\bibitem[Yamauchi et al.(2004)] {yamauchi2004} Yamauchi,~A., Nakai,~N., Sato,~N. \& Diamond, P., 2004, \pasj, 56, 605 

\bibitem[Zhang \& Ho(1995)] {zhang1995} Zhang,~Q., \& Ho,~P.~T.~P., 1995, \apj, 450L, 63

\end{thebibliography}
\end{document}